\documentclass[letterpaper,twocolumn,showpacs,prb,superscriptaddress,floatfix]{revtex4}
\usepackage[german,swedish,english]{babel}
\usepackage[dvips]{graphicx}
\usepackage{amssymb}
\usepackage{amsmath}
\usepackage{color}
\newcommand{\bm}{\mathbf}

\newcommand{\btau}{\mbox{\boldmath$\tau$}}

\begin{document}

\date{\today}

\title{Absence of low temperature anomaly in the Debye-Waller factor of solid $^4$He down to 140 mK}
\author{E. Blackburn}
\author{J. M. Goodkind}
\author{S. K. Sinha}
\affiliation{Physics Department, University of California, San Diego, 9500 Gilman Drive, La Jolla, California 92093\\}
\author{J. Hudis}
\author{C. Broholm}
\affiliation{Department of Physics and Astronomy, The Johns Hopkins University, Baltimore, Maryland 21218\\}
\author{J. van Duijn}
\affiliation{Departamento de Qu\'imica Inorg\'anica I, Facultad de Ciencias Qu\'imicas, Universidad Complutense de Madrid, 28040 Madrid, Spain}
\author{C. D. Frost}
\author{O. Kirichek}
\author{R. B. E. Down}
\affiliation{ISIS Facility, Rutherford-Appleton Laboratory, Chilton, Didcot, OX11 0QX, United Kingdom}

\begin{abstract}
The mean square atomic displacement in hcp-phase solid $^4$He has been measured in crystals with a molar volume of 21.3 cm$^3$.  It is temperature independent from 1 K to 140 mK, with no evidence for an anomaly in the vicinity of the proposed supersolid transition.  The mean square displacement is different for in-plane motions (0.122 $\pm$ 0.001 \AA $^{2}$) and out-of-plane motions (0.150 $\pm$ 0.001 \AA $^{2}$).  
\end{abstract}


\maketitle

\section{Introduction}

The recent experiments by Kim and Chan \cite{KC04a,KC04b} on hexagonal close packed (hcp) solid $^4$He have re-opened interest in the old question of supersolidity.  In the 1970s, several theorists considered whether the formation of a Bose-Einstein condensate was possible in the presence of a crystalline lattice \cite{AL69,Chester70,Leggett70} and Leggett \cite{Leggett70} predicted a signature in the rotational inertia.  Kim and Chan observed just such a shift, and this has subsequently been reproduced by several groups (e.g.~Ref.~\onlinecite{RR06}), although the origin and mechanism of this effect remain under discussion.  Experimental estimates of the onset of this phenomenon range from between $\sim$ 80 mK to 800 mK, dependent on the percentage of $^3$He present.\cite{KC04a}  An acoustic anomaly was observed by Ho \emph{et al.} \cite{HBG97} in the acoustic attenuation in solid $^4$He crystals, close to 200 mK. This was ascribed to a change in the behaviour of defects in the crystal, and one school of thought holds that defects may play an important role in supersolid formation.\cite{AL69}

To provide information on the the bulk behavior of solid $^4$He in this temperature range, we have carried out neutron diffraction experiments on solid $^4$He crystals to look for traces of a supersolid transition in the bulk material.  The neutron diffraction cross section is proportional to the square of the Fourier transform of the nuclear density distribution in space and time.  If the nuclei form a crystal lattice, then this Fourier transform yields Bragg peaks with an intensity that is proportional to the unit cell structure factor and the Debye-Waller factor, from which the mean square displacement of the $^4$He nuclei can be extracted.  Helium forms a quantum crystal and so the mean square displacement is greater than the Lindemann criterion due to large zero-point fluctuations.\cite{Werthamer69}  Depending on the model used to describe supersolidity, this quantity is liable to change in the supersolid state. As an aside, we note that even non-solid samples may have a periodic density distribution and hence Bragg peaks, and in this case, the effect on the mean square displacement would be even more significant.  

\section{Experimental Details}

The He crystals were prepared in a cylindrical stainless steel sample chamber (height 32.1 mm, diameter 33.5 mm) equipped with quartz transducers to monitor ultrasound propagation through the growing crystal.  $^4$He gas was supplied and pressurized through a stainless steel capillary located at the top of the sample chamber.  A mix containing a slightly elevated amount\cite{KC04a,HBG97} of $^3$He (40 ppm) was used.  From Ref.~\onlinecite{KC04a} this should place the start of the transition temperature at $\sim$ 550 mK, although the transition reported in Ref.~\onlinecite{HBG97} remains close to 200 mK. 

The sample chamber was cooled using an Oxford Instruments dilution refrigerator, wherein thermal contact to the mixing chamber was established using a Cu wire attached to the bottom of the sample chamber.  The sample chamber supported growth of multiple crystallites with different orientation as opposed to a single crystalline sample. 

The dilution refrigerator used for the measurements had difficulty cooling to 1 K when the filling capillary to the sample chamber was filled with liquid.  Therefore, the solid had to be formed at higher temperature and pressure.  The sample examined here was formed by starting at pressure $p$ = 45 bar and temperature $T$ $\sim$ 2 K.  Solid $^4$He was then grown using the `blocked capillary technique', in which a solid plug is formed in the capillary as the system is cooled so that the molar volume of the sample remains constant as it cools along the melting curve.  This means that our sample passed through the region of bcc phase on the melting curve,\cite{VF61} which presumably contributed to the formation of strained crystallites. 

The neutron data were collected using the MAPS instrument at the ISIS Facility, Rutherford Appleton Laboratory, Oxford.  MAPS is a time-of-flight (TOF) spectrometer equipped with approximately 16 m$^2$ of position sensitive detectors located 6 m from the sample.  The instrument was operated in TOF Laue mode with a white pulsed incident beam and the sample fixed in place.   Diffraction patterns were obtained over the temperature range 140 mK - 800 mK.  At the end of the experiment the sample can was evacuated and an empty can measurement was acquired at 1 K.  A standard vanadium sample was also measured to obtain information on the wavelength dependent detector efficiency.

\section{Data Analysis}

To assess the mean square displacement, multiple peaks from the same crystallite are required, preferably over a large range of momentum transfer.  Due to the presence of multiple crystallites in the sample can, it was impractical to identify peaks from the same crystallite at different locations on the detector.  However, because we used a pulsed white incident beam, higher-order reflections from the same crystallite appeared in the same detector pixels, separated by the recorded time-of-flight.  For these higher-order reflections to be visible, the principal reflection had to be sufficiently strong.  45 reflections were indexed on the basis of their $d$-spacings, assuming an hcp structure.  Peaks of the types (002), (100), (101) and (102) were readily identified, but only peaks of the type (002) and (100) were strong enough for the corresponding higher-order reflections (004) and (200) to be visible at half the TOF. 

The molar volume was determined using the range of plane spacings obtained at this point, even though the peaks were not from the same crystallites. We found $a$ = 3.68(1) \AA~and $c$ = 6.03(1) \AA, corresponding to a molar volume of 21.3(1) cm$^3$. The $c/a$ ratio is 1.638(5) as compared with the ideal hcp ratio of 1.633.  This is consistent with expectations from the temperature and pressure during the crystal growth process.\cite{Swenson50}  

Once these peaks were identified, the TOF spectra in the corresponding pixels were normalized to the incident flux on the sample and the empty can measurement subtracted.  At small momentum transfers, where the majority of scattering from the can and cryostat was found, a small amount of background scattering remained after this subtraction.  An additional subtraction was made when integrating over reflections by taking a slice from the same pixel over a $d$-spacing range close to that of interest.  At higher momentum transfers, the initial subtraction is sufficient.   

The integrated intensities were then corrected for (i) angular effects, (ii) detector efficiency and (iii) the wavelength dependence of the incident beam flux.  To establish the first correction factor, consider a diffraction peak with a scattering angle 2$\theta_B$ acquired on a 2D plate a distance $Z$ (= 6.0031 m) from the sample.  The centre of the detector plate coincides with the centre of an orthonormal coordinate system where $\bf{\hat{x}}$ and $\bf{\hat{y}}$ are perpendicular directions in the plane of the detector, and $\bf{\hat{z}}$ is perpendicular to the plate.  The reflection is observed at point ($\bf{x}$,$\bf{y}$) on the plate, and distance from the centre of the sample to this point is $R = ( \bm{x} ^2 + \bm{y} ^2 + Z^2 )^{1/2}$.  The direction of the scattered beam can be expressed as a function of two angles, corresponding to the latitude and longitude (labelled $\psi$ and $\phi$).  Using these angles, the position of the Bragg peak is $R(\cos{\psi} \sin{\phi}, - \sin{\psi}, \cos{\psi} \cos{\phi})$, keeping the $(\bf{\hat{x}},\bf{\hat{y}},\bf{\hat{z}})$ coordinate system.  The scattering vector $\bm{Q}$ is $(\frac{k\bm{x}}{R},\frac{k\bm{y}}{R},k(\frac{\bm{z}}{R} - 1))$ and so $\bm{Q}^2 = 2 k^2 (1 - \cos{\psi}\cos{\phi}) = 4 k^2 \sin^2{\theta_B}$, where $k$ is the magnitude of the wavevector.  

The number of neutrons counted in the Bragg peak is 
\begin{equation}
I = \frac{\mathrm{d}\sigma}{\mathrm{d}\Omega} . \Delta\Omega . \Phi
\end{equation}
where $\Phi$ is the incident flux on the sample (neutrons / unit area / unit time), and $\Delta \Omega$ is the solid angle subtended by the sample. We convert this to an integral form; the solid angle can be written as $(d\bm{S}\cdot\bm{R}) / R^3 = \mathrm{d}\bm{x}\mathrm{d}\bm{y} Z / R^3$.   The incident flux can be written as a function of the incident wavelength - any variation from unity here is accounted for when the vanadium scans are considered.
\begin{equation}
\Phi = I(\lambda) \mathrm{d}\lambda = \frac{\lambda^2 I(\lambda)}{2\pi} \mathrm{d}k
\end{equation}
The neutron scattering cross-section is
\begin{equation}
\frac{\mathrm{d}\sigma}{\mathrm{d}\Omega} = (2\pi)^3 N \vert F(\btau) \vert^2 \delta(\bm{Q} - \btau)
\end{equation}
where $N$ is the number of unit cells divided by the volume of a unit cell, and is particular to each crystallite, and $\vert F(\btau) \vert^2$ is the structure factor of the Bragg peak.  We therefore arrive at the following expression for the intensity of the Bragg peak at reciprocal lattice vector $\btau$.
\begin{equation}
I_{B} = (2\pi)^3 N \vert F(\btau) \vert^2 \frac{\lambda^2 I(\lambda)}{2\pi} \frac{Z}{R^3} \int \int \int \delta(\bm{Q} - \btau) \; \mathrm{d}x \mathrm{d}y \mathrm{d}k
\end{equation}
Integrating over momentum space, the Bragg peak intensity is given by
\begin{equation}
I_{B} = 2N \vert F(\btau) \vert^2 d_{hkl}^2 \lambda^2 I(\lambda).
\label{eq:Imeas}
\end{equation}
This expression applies equally to all of the detector banks.  The value of $d_{hkl}$ for the main peaks was obtained by fitting the peak profile as a Gaussian and taking the central value.  This value was then halved for the relevant higher-order peak.  To correct for the other factors, the vanadium scan was used.  

Vanadium is a uniform scatterer, and so 
\begin{eqnarray}
I_{\mathrm{vana}} & = & \frac{\sigma_v}{4\pi} I(\lambda) \Delta \lambda \Delta A_d \frac{Z}{R^3} \nonumber \\
& = & \frac{\sigma_v}{4\pi} \frac{\Delta A_d}{Z^2} \cos^3{2\theta_B} \sin{\theta_B} I(\lambda) \Delta d
\label{eq:Ivana}
\end{eqnarray}
where $\sigma_v$ = $5.08 \cdot 10^{-28}$ m$^2$ is the incoherent cross-section for vanadium,\cite{NDB} $Z$ = 6.0031 m, and $\Delta A_d = N_p \Delta x \Delta y$ is the relevant area of the detector ($N_p$ is the number of pixels).  For one pixel $\Delta x$ = 0.025 m (detector tube diameter) and $\Delta y$ = 0.015 m (vertically resolvable distance).  
Eq.~\ref{eq:Ivana} allows $I(\lambda)$ to be extracted from the white beam vanadaium diffraction data.  

In the harmonic approximation, the structure factor $\vert F(\btau) \vert^2$ is related to the mean-square displacement as follows
\begin{equation}
F(\btau) = b_{He} \sum_i \exp(i \btau \cdot \bm{d}_i) \exp(-W_{\bm{d}_i}(\btau)).
\label{eq:Ftau}
\end{equation}
Here $\bm{d}_i$ is the position of the i$^{\mathrm{th}}$ atom in the unit cell, $b_{He}$ = 3.26$\cdot 10^{-15}$ m is the scattering length of $^{4}$He,\cite{NDB} $W_{\bm{d}_i}(\btau) = \langle (\btau\cdot \bm{u})^2 \rangle$, and $\bf{u}$ is the displacement from the average periodic lattice for a specific nucleus.\cite{Squires}  For the He hcp crystal, the lowest order expansion of the mean square displacement can be broken down into in-plane and out-of-plane ($\parallel$ to $c$) components, as follows
\begin{equation}
\langle (\btau \cdot \bm{u})^2 \rangle = \tau^2 \lbrack \langle u_{\parallel}^2 \rangle (\hat{\btau}\cdot\hat{c})^2) + \langle u_{\perp}^2 \rangle (1 - (\hat{\btau}\cdot\hat{c})^2)) \rbrack.
\end{equation}
Previous studies have found this expression to be quite adequate for describing the observed data.\cite{SKK78,VS03}

\begin{figure}[tb]
\begin{center}
\includegraphics[width=0.4\textwidth]{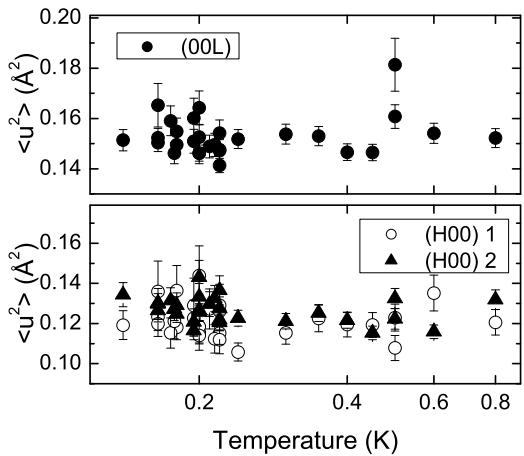}
\end{center}
\caption{The mean square displacement, $\langle u^2 \rangle$, of $^4$He hcp crystal, as measured for the three selected peaks as a function of temperature.}
\label{fig:u2plot}
\end{figure}

Hcp He has two atoms per unit cell, and due to symmetry equivalence $W_{\bm{d}_i}$ has no site dependence, and so can be pulled out of the summation over $i$.  After making these corrections, one finds that
\begin{equation}
\vert \exp{-(W_{\bm{d}_i})} \vert^2 = \frac{I_{\mathrm{corr}}}{N}
\end{equation}
where $I_{\mathrm{corr}} = C (I_{B} / I_{\mathrm{vana}})$ is the corrected intensity of each peak, and $C$ is the total correction factor, obtained from Eqs.~\ref{eq:Imeas}, \ref{eq:Ivana} and \ref{eq:Ftau}.

We therefore have
\begin{equation}
\ln{I_{\mathrm{corr}}} = \ln{N} - 2 Q^2 \lbrack \langle u_{\parallel}^2 \rangle (\hat{\bm{Q}}\cdot\hat{c})^2) + \langle u_{\perp}^2 \rangle (1 - (\hat{\bm{Q}}\cdot\hat{c})^2)) \rbrack
\end{equation}
and so the mean square displacement can now be determined from a simple linear fit of intensity vs $Q^2$.

The level of thermal diffuse scattering, calculated following the method of Popa and Willis \cite{PW97} for TOF neutron diffractometry, using the phonon velocities measured by Minkiewicz \textit{et al.},\cite{MKLN+68} was found to be negligible.

\section{Results}

\begin{figure}
\begin{center}
\includegraphics[width=0.4\textwidth]{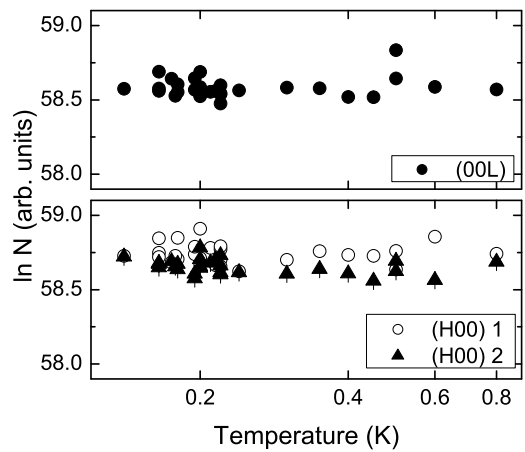}
\end{center}
\caption{The crystallite dependent quantity $\ln N$, as measured from $^4$He hcp crystal for the three selected peaks as a function of temperature.}
\label{fig:lnNplot}
\end{figure}

Three peaks met the conditions established above, and values for $\langle u^2 \rangle$ were obtained at each temperature considered.  Figures \ref{fig:u2plot} and \ref{fig:lnNplot} show the temperature dependence of $\langle u^2 \rangle$ and $\ln N$.  The data are tabulated in the appendix.  No thermal anomaly indicative of a phase transition is observed, consistent with recent path-integral computations that indicate that bulk solid He crystals cannot be supersolid.\cite{CB04}  Table \ref{tab:peak_results} averages over all of the temperatures assuming no temperature dependence.  $\langle u^2 \rangle$ is higher for the (00L) peaks than for the (H00) peaks indicating some anisotropy.   The similar values of $\ln N$ indicate similar crystallite sizes.  Consideration of the actual peak positions on the detector indicate that the second and third peaks are (100) type peaks that could be from the same crystallite, while the (002) type peak is from a different crystallite.  

\begin{figure}
\begin{center}
\includegraphics[width=0.4\textwidth]{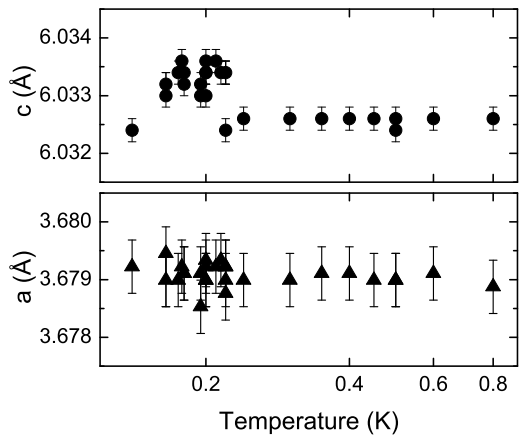}
\end{center}
\caption{The temperature dependence of the $c$ and $a$ lattice parameters from two of the crystallites studied in the text.}
\label{fig:lattice_param}
\end{figure}

Figure \ref{fig:lattice_param} shows the temperature dependence of the lattice parameter $a$ [$c$] as measured from the (100) [(002)] reflection.  Although the lattice parameters for a given crystallite remained fairly stable as a function of temperature, due to different annealing conditions and difficulties with temperature stability between different runs, there was some change in the values obtained.  This is most clearly obvious in the upper panel of Figure \ref{fig:lattice_param}, where two regimes are visible.  The upper set of points, which differ from the rest by 1 part in 6000, correspond to conditions where the stress on the crystallite was believed to be slightly different to that present for the other measured points.  These changes had little impact on the observed mean square displacement, indicating that we are observing the (temperature- and volume-independent) zero-point motion here.  This is in agreement with the observations made at a lower molar volume by Adams \textit{et al.}~\cite{AMKD07} from 0.07 - 0.4 K.

\section{Discussion}

Table \ref{tab:literature_values} lists published data for $\langle u^2 \rangle$ at different molar volumes.  The form for the Debye-Waller factor differs in some of these references, and so all values have been converted to the definition of $\langle u^2 \rangle$ given above.  These values compare well with theoretically calculated numbers; the value for a crystal of molar volume 15.72 cm$^3$ is reproduced very well by path-integral Monte Carlo techniques,\cite{DC00} and the value at 20.9 cm$^3$ is close to that found by Whitlock \emph{et al.} using the Green's function Monte Carlo method.\cite{WCCK79}  Some estimates were obtained from inelastic neutron scattering studies of the phonons, but the validity of these estimates have been called into doubt by Burns and Isaacs,\cite{BI97} who claim that the phonon data suffer from significant multiple scattering and are not included. 

We have observed a certain amount of anisotropy between in-plane and out-of-plane displacements.  The in-plane value is 20 \% smaller than the out-of-plane value, indicating that atomic motion is easier out-of-plane.  This is not unexpected for an hcp structure, particularly as the close packing is not perfect, and the qualitative behavior agrees with that calculated by Reese \textit{et al.}~\cite{RSBT71} from phonon frequencies.  The previous x-ray study by Venkataraman and Simmons \cite{VS03} did not report any anisotropy in the hcp phase, but the reflections studied are not noted.  Venkataraman and Simmons noted a decrease in $\langle u^2 \rangle$ as the temperature dropped in the fcc solid phase.  We see no temperature dependence, which implies that we are looking at quantum fluctuations. 

\begin{center}
\begin{table}
\begin{tabular}{|c|c|c|c|c|c|}
\hline
& Peak & $d$-spacing (\AA) & 2$\theta_B$ ($^{\circ}$) & $\langle u^2 \rangle$ (\AA $^2$)& $\ln{N}$ \\
\hline
1 & 002 & 3.016 & 20.6 & 0.150 $\pm$ 0.001 &  58.560 $\pm$ 0.002 \\
2 & 100 & 3.188 & 18.8 & 0.118 $\pm$ 0.001 &  58.716 $\pm$ 0.003 \\ 
3 & 100 & 3.186 & 19.7 & 0.125 $\pm$ 0.001 &  58.636 $\pm$ 0.002 \\
\hline
\end{tabular}
\caption{Characteristic properties of the peaks examined on MAPS.}
\label{tab:peak_results}
\end{table}
\end{center}

%\subsection{Comparison with the literature}

\begin{center}
\begin{table}
\begin{tabular}{|c|c|c|c|l|}
\hline
Molar & Temp. &$\langle u^2 \rangle$ & Peak & Reference\\
volume (cm$^3$) & (K) & (\AA $^2$) & type & \\
\hline 
 11.01 & 15 & 0.0593(1) & mixed & Ref.~\onlinecite{VS03} (x-ray)  \\
 12.06 & 5.8 & 0.0466(3)  & mixed & Ref.~\onlinecite{SKK78} (neutron) \\  
 12.12 & 14.8 & 0.0563(14) & mixed & Ref.~\onlinecite{Arms99} (x-ray)  \\
 12.13 & 14.8 & 0.0513(10) & mixed & Ref.~\onlinecite{Arms99} (x-ray)  \\               
 15.72 & 5.8 & 0.0861(9)  & mixed & Ref.~\onlinecite{SKK78} (neutron) \\
 20.9 & 0.7 & 0.1537(7)   & (00L) & Ref.~\onlinecite{BI97} (x-ray) \\
 21.3 & $<$1 & 0.150(1)    & (00L) & this work (neutron) \\                 
 21.3 & $<$1 & 0.122(1)    & (H00) & this work (neutron) \\
\hline
\end{tabular}
\caption{Published values for $\langle u^2 \rangle$ in solid $^4$He obtained by x-ray and neutron diffraction.  In all cases, the harmonic approximation is assumed and for the mixed peak data, no distinction between in-plane and out-of-plane displacements were made.}
\label{tab:literature_values}
\end{table}
\end{center}

\begin{figure}
\begin{center}
\includegraphics[width=0.4\textwidth]{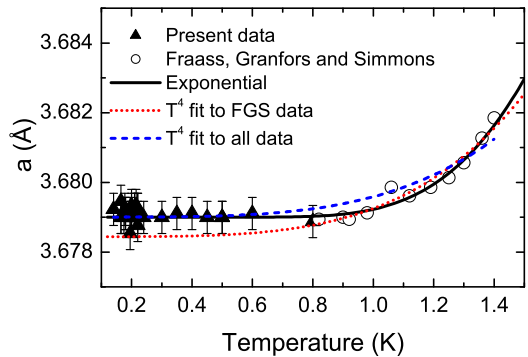}
\end{center}
\caption{The temperature dependence of the $a$ lattice parameter as measured here (closed points) and as calculated from Fig.~2 of Ref.~\onlinecite{FGS89} (open points), assuming that the temperature independent part of $a$ is the same for both data sets.  The lines are based on the different models for vacancy concentration: the solid line is the exponential from the thermal activation model, with an activation energy of 8.6 K; the broken lines are fits to the model proposed in Ref.~\onlinecite{ABH05} - a $T^4$ dependence.  The dotted line is a fit to the data from Ref.~\onlinecite{FGS89} only and the dashed line includes the data measured in this paper.}
\label{fig:vacancy_fig}
\end{figure}

The lattice parameter measurements allow us to comment on the model proposed by Anderson \textit{et al.}~\cite{ABH05}, that posits a $T^4$ dependence for the vacancy concentration, as opposed to a classical thermally activated model.  Fraass \textit{et al.}~\cite{FGS89} used the change in lattice parameter with temperature as a measure of the vacancy concentration.  If the actual mass density of He atoms is constant as a function of temperature, this approach is valid, and the caveats associated with this analysis are outlined in Ref.~\onlinecite{FGS89}. From Fraass \textit{et al.}~'s data, the two models could not be separated, but with the extension to temperatures below 0.8 K provided here, the thermally activated model is favoured (see Fig.~\ref{fig:vacancy_fig}), and fitting the complete data set gives an formation energy of (8.6 $\pm$ 0.1) K.   We note that the samples compared here had slightly different molar volumes, and that in our sample chamber, there were several crystallites and so the volume of a given crystallite may have varied as a function of temperature.

As a complement to this, we note that acoustic attenuation data collected by one of us \cite{LG90} provide an additional confirmation of this exponential dependence, as the acoustic attenuation is determined by the combination of phonon and delocalized vacancy contributions.  

\section{Conclusions}

The principal result of this paper is that the mean square displacement does not change over the temperature range measured here.  Within the precision of these data, there is no apparent temperature dependence at all over the temperature range studied, i.e.~there is no indication that the supersolid transition, if it exists, affects the crystalline lattice or zero-point fluctuations.  We have also observed a measure of anisotropy between in-plane and out-of-plane motions.

We thank the staff at the ISIS Facility, Rutherford Appleton Laboratory, Oxford, for their assistance with the low-temperature and pressure equipment needed.

\appendix*

\section{Data Tables}
\begin{table}
\begin{tabular}{c|c|c|c|c|c|c}
\multicolumn{1}{c}{ } & \multicolumn{2}{c}{00L} & \multicolumn{2}{c}{H00a} & \multicolumn{2}{c}{H00b} \\
\hline
Temp. & $\langle u^2 \rangle$& error & $\langle u^2 \rangle$ & error & $\langle u^2 \rangle$ & error \\
(K) & (\AA $^2$) & (\AA $^2$) & (\AA $^2$) & (\AA $^2$) & (\AA $^2$) & (\AA $^2$) \\
\hline \hline
0.140 & 0.151 & 0.004 & 0.119 & 0.007 & 0.134 & 0.006 \\
0.165 & 0.165 & 0.009 & 0.136 & 0.015 & 0.130 & 0.008 \\
0.165 & 0.152 & 0.004 & 0.120 & 0.006 & 0.130 & 0.005 \\
0.165 & 0.150 & 0.004 & 0.123 & 0.007 & 0.126 & 0.004 \\
0.175 & 0.159 & 0.006 & 0.115 & 0.008 & 0.131 & 0.006 \\
0.178 & 0.146 & 0.004 & 0.121 & 0.009 & 0.127 & 0.006 \\
0.180 & 0.155 & 0.005 & 0.136 & 0.012 & 0.129 & 0.006 \\
0.180 & 0.150 & 0.003 & 0.118 & 0.006 & 0.125 & 0.004 \\
0.195 & 0.151 & 0.005 & 0.123 & 0.009 & 0.116 & 0.005 \\
0.195 & 0.160 & 0.008 & 0.129 & 0.014 & 0.121 & 0.007 \\
0.200 & 0.153 & 0.005 & 0.114 & 0.007 & 0.133 & 0.007 \\
0.200 & 0.164 & 0.007 & 0.118 & 0.008 & 0.127 & 0.006 \\
0.200 & 0.146 & 0.004 & 0.144 & 0.015 & 0.143 & 0.008 \\
0.200 & 0.146 & 0.003 & 0.116 & 0.006 & 0.126 & 0.004 \\
0.210 & 0.149 & 0.005 & 0.127 & 0.010 & 0.130 & 0.006 \\
0.215 & 0.150 & 0.005 & 0.112 & 0.007 & 0.132 & 0.007 \\
0.220 & 0.147 & 0.004 & 0.129 & 0.010 & 0.122 & 0.005 \\
0.220 & 0.148 & 0.004 & 0.126 & 0.010 & 0.127 & 0.006 \\
0.220 & 0.154 & 0.005 & 0.112 & 0.007 & 0.137 & 0.007 \\
0.220 & 0.141 & 0.003 & 0.115 & 0.006 & 0.121 & 0.004 \\
0.240 & 0.152 & 0.004 & 0.106 & 0.005 & 0.123 & 0.004 \\
0.300 & 0.154 & 0.004 & 0.115 & 0.006 & 0.121 & 0.004 \\
0.350 & 0.153 & 0.004 & 0.123 & 0.007 & 0.125 & 0.004 \\
0.400 & 0.147 & 0.003 & 0.120 & 0.006 & 0.122 & 0.004 \\
0.450 & 0.147 & 0.003 & 0.119 & 0.006 & 0.115 & 0.003 \\
0.500 & 0.161 & 0.005 & 0.123 & 0.007 & 0.132 & 0.005 \\
0.500 & 0.181 & 0.011 & 0.108 & 0.006 & 0.122 & 0.005 \\
0.600 & 0.154 & 0.004 & 0.135 & 0.009 & 0.116 & 0.003 \\
0.800 & 0.152 & 0.004 & 0.121 & 0.006 & 0.132 & 0.005 \\
\hline
\end{tabular}
\caption{The mean square displacement $\langle u^2 \rangle$ and associated error for all temperatures measured, as measured for three selected Bragg peaks from hcp crystals of $^4$He.}
\label{tab:u2data}
\end{table}

\begin{table}
\begin{tabular}{c|c|c|c|c|c|c}
\multicolumn{1}{c}{ } & \multicolumn{2}{c}{00L} & \multicolumn{2}{c}{H00a} & \multicolumn{2}{c}{H00b} \\
\hline
Temp. & lnN & error & lnN & error & lnN & error\\
(K) & (\AA $^2$) & (\AA $^2$) & (\AA $^2$) & (\AA $^2$) & (\AA $^2$) & (\AA $^2$) \\
\hline \hline
0.140 & 58.574 & 0.009 & 58.727 & 0.015 & 58.721 & 0.013 \\
0.165 & 58.834 & 0.021 & 58.636 & 0.013 & 58.623 & 0.011 \\
0.165 & 58.560 & 0.008 & 58.748 & 0.015 & 58.649 & 0.009 \\
0.165 & 58.577 & 0.008 & 58.720 & 0.013 & 58.679 & 0.010 \\
0.175 & 58.689 & 0.018 & 58.845 & 0.031 & 58.673 & 0.016 \\
0.178 & 58.552 & 0.007 & 58.705 & 0.013 & 58.636 & 0.009 \\
0.180 & 58.524 & 0.007 & 58.691 & 0.012 & 58.645 & 0.009 \\
0.180 & 58.476 & 0.006 & 58.701 & 0.012 & 58.601 & 0.008 \\
0.195 & 58.564 & 0.008 & 58.624 & 0.010 & 58.616 & 0.009 \\
0.195 & 58.583 & 0.008 & 58.701 & 0.012 & 58.606 & 0.008 \\
0.200 & 58.578 & 0.008 & 58.758 & 0.014 & 58.638 & 0.009 \\
0.200 & 58.520 & 0.007 & 58.733 & 0.013 & 58.607 & 0.009 \\
0.200 & 58.518 & 0.007 & 58.728 & 0.013 & 58.559 & 0.008 \\
0.200 & 58.644 & 0.010 & 58.760 & 0.014 & 58.692 & 0.011 \\
0.210 & 58.587 & 0.008 & 58.857 & 0.018 & 58.564 & 0.008 \\
0.215 & 58.570 & 0.008 & 58.742 & 0.014 & 58.686 & 0.010 \\
0.220 & 58.532 & 0.009 & 58.910 & 0.030 & 58.782 & 0.017 \\
0.220 & 58.688 & 0.014 & 58.712 & 0.017 & 58.657 & 0.012 \\
0.220 & 58.555 & 0.009 & 58.781 & 0.020 & 58.680 & 0.013 \\
0.220 & 58.560 & 0.010 & 58.664 & 0.015 & 58.700 & 0.014 \\
0.240 & 58.543 & 0.009 & 58.772 & 0.020 & 58.661 & 0.012 \\
0.300 & 58.538 & 0.009 & 58.791 & 0.021 & 58.615 & 0.011 \\
0.350 & 58.598 & 0.011 & 58.657 & 0.015 & 58.731 & 0.015 \\
0.400 & 58.585 & 0.010 & 58.672 & 0.015 & 58.703 & 0.014 \\
0.450 & 58.642 & 0.012 & 58.684 & 0.016 & 58.690 & 0.013 \\
0.500 & 58.527 & 0.009 & 58.728 & 0.018 & 58.655 & 0.012 \\
0.500 & 58.604 & 0.011 & 58.849 & 0.025 & 58.673 & 0.013 \\
0.600 & 58.569 & 0.010 & 58.739 & 0.019 & 58.574 & 0.010 \\
0.800 & 58.646 & 0.016 & 58.788 & 0.028 & 58.606 & 0.014 \\
\hline
\end{tabular}
\caption{The crystallite dependent quantity ln N and associated error for all temperatures measured, as obtained from $^4$He hcp crystals for three selected peaks.}
\label{tab:lnNdata}
\end{table}

\end{document}